\documentclass[aps,pra,twocolumn,groupedaddress,showpacs]{revtex4-1}
\usepackage{graphicx}
\usepackage{amsmath}
\usepackage{nicefrac}
\begin{document}

%user-defined macros
\newcommand{\barb}{\overline{b}}
\newcommand{\barbh}{\overline{b}^{(0)}}
\newcommand{\barpsi}{\overline{\psi}}
\newcommand{\barphi}{\overline{\phi}}
\newcommand{\barbeta}{\overline{\beta}}
\newcommand{\dkap}{\frac{d\kappa}{\sqrt{2\pi}}}
\newcommand{\dkapp}{\frac{d\kappa'}{\sqrt{2\pi}}}
\newcommand{\barGamma}{\overline{\Gamma}}

%\preprint{pra}

\title{Spontaneous four-wave mixing in lossy microring resonators}

\author{Z. Vernon}
\email{zachary.vernon@utoronto.ca}
\author{J.E. Sipe}
%\email[]{Your e-mail address}
%\homepage[]{Your web page}
%\thanks{}
%\altaffiliation{}
\affiliation{Department of Physics, University of Toronto, 60 St. George Street, Toronto, Ontario, Canada, M5S 1A7}

\date{\today}

\begin{abstract}
We develop a general Hamiltonian treatment of spontaneous four-wave mixing in a microring resonator side-coupled to a channel waveguide. The effect of scattering losses in the ring is included, as well as parasitic nonlinear effects including self- and cross-phase modulation. A procedure for computing the output of such a system for arbitrary parameters and pump states is presented. For the limit of weak pumping an expression for the joint spectral intensity of generated photon pairs, as well as the singles-to-coincidences ratio, is derived.
\end{abstract}


\maketitle

\section{Introduction} \label{sec:intro}
Quantum states of light, including single photons and entangled photon pairs, are a critical resource in quantum information processing. While traditional techniques for generating entangled photon pairs have relied on bulk nonlinear optical materials and waveguides \cite{Takesue2004,Fiorentino2007}, a less cumbersome micro-scale implementation is needed for an integrated optics setting. Designing such ``on-chip" integrated optical components is likely a necessary step in the physical realization of optical quantum information processing. Recent advances in the fabrication of resonant optical microstructures \cite{Xu2008,Xia2007} have yielded promising chip-based candidates for entangled photon pair generation. One such implementation involves a micron-scale optical ring resonator side-coupled to a channel waveguide. By pumping the ring via the side channel, the nonlinear optical response in the ring may give rise to entangled photon pairs generated by spontaneous parametric down-conversion (SPDC) or spontaneous four-wave mixing (SFWM). By operating at wavelengths close to the ring resonances, dramatic enhancement of nonlinear conversion efficiencies has been demonstrated with peak pump powers of only a few mW \cite{Ferrera2008, Ferrera2009,Turner2008}. Photon pair generation from both SPDC and SFWM in silicon ring resonators \cite{Azzini2012, Azzini2012a, Clemmen2009, Engin2013,Turner2008, Grassani2015, Wakabayashi2015}, classical four-wave mixing \cite{Azzini2012a,Ferrera2008,Ferrera2009} and optical parametric oscillation \cite{Levy2010} in silicon nitride and doped silica glass rings, as well as frequency comb generation in silicon nitride and aluminum nitride rings \cite{Okawachi2011,Jung2013} have recently been experimentally demonstrated. There has been some theoretical study of SFWM and SPDC in these systems for the case of a continuous wave pump \cite{Chen2011, Camacho2012}, and perturbative calculations have been performed which demonstrate how controlling the pump pulse duration affects spectral correlations in the generated entangled photon pairs \cite{Helt2010,Yang2007,Yang2007a,Helt2012,Liscidini2012}, but so far these have neglected the quantum effects of losses on the generated pairs due to scattering of photons in the ring. Parasitic nonlinear effects such as self- and cross-phase modulation (SPM and XPM, respectively), which may become important for strong pumps, as well as other strong-pumping corrections have not been fully included in these theoretical treatments of photon pair generation.

In this paper we present a Hamiltonian treatment of third-order nonlinear optical processes in such a system, focusing on SFWM while accounting for SPM of the pump field, as well as XPM between the pump, signal and idler fields. We include scattering losses in the ring, and show that they may be modelled as an effective extra channel into which ring modes can emit. Probabilities corresponding to single-photon detection events wherein a photon's pair partner is lost to scattering, which are not directly calculable in the absence of a formal loss mechanism, can then be studied.

In the next section we begin by laying out the Hamiltonian for the full ring-channel system and bath. Heisenberg equations of motion for each field in the channel and ring are developed in section \ref{sec:eqns_of_motion}, and the notion of incoming and outgoing fields is formalized. In section \ref{sec:solving_eqns_of_motion} a strategy is developed to solve these equations for arbitrary parameters and pump pulses. Within our treatment the outgoing and incoming field operators are related in the frequency domain by spectral response functions, determined by solving equations of motion for the ring operators alone. A specific perturbative example for weak pumps is worked out in section \ref{sec:perturbative}, the results of which we then use to calculate the joint spectral intensity of generated photon pairs, as well as the singles-to-coincidences ratio.

\section{Model Hamiltonian}
\begin{figure}
\includegraphics[width=1.0\columnwidth]{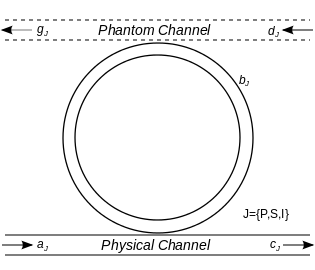}
\caption{Ring and channel microstructure geometry with labelled incoming, outgoing and ring modes. Upper ``phantom channel" represents scattering loss mechanism in the ring. \label{fig:ringchannel}}
\end{figure}

We begin with a Hamiltonian for the channel and ring system as shown in Fig. \ref{fig:ringchannel}. We take the channel to extend from $z=-\infty$ to $z=+\infty$ and to accommodate three fields of interest: one pump with reference frequency $\omega_P$, as well as a signal and idler field with reference frequencies $\omega_S$ and $\omega_I$, respectively. Each of these fields couple linearly to the ring, with the coupling assumed to take place at a single point in space, $z=0$. We further assume the ring geometry and field frequencies have been chosen such that the signal, idler and pump frequencies lie within narrow, well separated resonances. To begin with we focus on an individual pair of signal-idler fields, and discuss the case of multiple sets of signal-idler fields in later sections.

Extending the Hamiltonian considered earlier \cite{Helt2010}, within our model the fully quantum-mechanical Hamiltonian can be split into four parts,
\begin{eqnarray}\label{hamiltonian_split_up}
H = H_{channel} + H_{ring} + H_{coupling} + H_{bath},
\end{eqnarray}
wherein $H_{channel}$ refers to the free Hamiltonian for the channel fields, $H_{ring}$ to the ring, $H_{coupling}$ to the linear channel-ring coupling, and $H_{bath}$ to the modes into which photons can be lost, as well as their couplings to the ring modes. We then have for the channel Hamiltonian
\begin{eqnarray}\label{H_channel}
 H_{channel} = && \sum_J \Biggl[ \hbar\omega_J\int dz \psi_J^\dagger(z)\psi_J(z) \nonumber\\
&& + \frac{i\hbar v_J}{2}\int dz\left(\frac{d\psi_J^\dagger(z)}{dz}\psi_J(z) - \mathrm{h.c.}\right)\Biggl].
\end{eqnarray}
In this expression (and elsewhere in this work) the index $J$ ranges over the mode labels $\lbrace P, S, I\rbrace$. The channel fields $\psi_J(z)$ can be decomposed into constituent modes with lowering operators $A_J(\kappa)$ according to
\begin{eqnarray}\label{original_field_defn}
\psi_J(z) = \int \frac{d\kappa}{\sqrt{2\pi}}A_J(\kappa)e^{i(\kappa - \kappa_{J})z},
\end{eqnarray}
where $\kappa_{J}$ is the reference wavevector corresponding to frequency $\omega_J$ for the field $\psi_J(z)$, and is subtracted in the exponential to lead to a $\psi_J(z)$ that is slowly varying in space. If the integrals in (\ref{original_field_defn}) ranged from $-\infty$ to $\infty$, the channel operators $\psi_J(z)$ would satisfy
\begin{eqnarray}\label{commutation_relations}
\left[\psi_J(z),\psi_{J'}(z')\right] &&= 0, \nonumber\\
\left[\psi_J(z),\psi_{J'}^\dagger(z')\right] &&= \delta(z-z')\delta_{JJ'}.
\end{eqnarray}
While the integrals in (\ref{original_field_defn}) do not span the whole range of $\kappa$, both because of mode cut-offs and because the same transverse mode of the waveguide might be broken into different spectral ranges to identify different fields $\psi_J(z)$, we assume we work with pulses sufficiently long that (\ref{commutation_relations}) is a good approximation.
The coefficients $v_J$ in $H_{channel}$ refer to the propagation speeds for each field $J$, accounting for group velocity dispersion between the different channel fields. While our model neglects group velocity dispersion between modes within a single channel field, our treatment can easily be generalized to include arbitrary dispersion. 

Whereas in our model each field in the the channel consists of a continuum of modes, we approximate the fields in the ring as isolated individual modes, which is valid provided the ring resonances are sufficiently narrow \cite{Liscidini2012}. The ring Hamiltonian can then be written in terms of discrete mode annihilation operators $b_J$,
\begin{eqnarray}
H_{ring} = \sum_J\hbar\omega_J b_J^\dagger b_J + H_{NL}.
\end{eqnarray}
The latter term in the above expression is responsible for all the nonlinearity in this system. Since the pump field intensity in the ring resonator will be much larger than that in the channel, we assume that $H_{NL}$ contains only ring operators. In this work we focus on effects arising from the third-order nonlinear susceptibility $\chi^{(3)}$, taking
\begin{eqnarray}
H_{NL} =&& \left(\hbar\lambda b_Pb_Pb_S^\dagger b_I^\dagger + \mathrm{h.c.}\right) + \hbar\eta b_P^\dagger b_P^\dagger b_P b_P \nonumber \\
&& + \hbar\zeta\left(b_S^\dagger b_P^\dagger b_S b_P + b_I^\dagger b_P^\dagger b_I b_P\right).
\end{eqnarray}
The first term in this expression leads to photon pair generation: two pump photons may be converted into a signal and idler photon pair. The second represents SPM of the pump, while the latter two correspond to XPM between the pump and the signal and idler modes. SPM of the signal and idler modes is neglected, as the number of photons in those modes will be small compared to that in the pump mode. The nonlinear coupling coefficients $\lambda, \eta$ and $\zeta$ are in general not independent, for they arise from the same underlying optical nonlinearity, and can be calculated from the mode fields \cite{Sipe2004} in the ring. A simple estimate gives $\hbar\lambda\approx 3(\hbar\omega_P)^2\chi^{(3)}/(4\epsilon_0 n^4 \Omega_{ring})$, with $\eta=\lambda/2$ and $\zeta=2\lambda$, where $n$ is the refractive index of the ring and $\Omega_{ring}$ the volume of the ring mode, and $\epsilon_0$ the permittivity of free space, assuming only one component of the nonlinear susceptibility $\chi^{(3)}$ is accessed uniformly in the ring. For ease of exposition we concentrate on degenerate SFWM, though the methods presented here may easily be generalized to include multiple pump fields, as well as effects arising from other nonlinearities, including SPDC. Our formalism also readily encompasses classical four-wave mixing if the input state in each mode is treated as a coherent state.

We adopt a point-coupling model for the interaction between the channel and ring. This approximation is justified provided the spatial extent of the interaction region is short compared to all relevant wavelengths \cite{Heebner2008}. Our ring-channel coupling Hamiltonian is then
\begin{eqnarray}
H_{coupling} = \sum_J\left(\hbar\gamma_Jb_J^\dagger\psi_J(0) + \mathrm{h.c.}\right),
\end{eqnarray}
in which we have introduced the ring-channel coupling coefficients $\gamma_J$. These constants can be related to the usual self- and cross-coupling coefficients \cite{Heebner2008} used to characterize microring resonators \cite{Chak2006}.

In an experimental setting, loss due to scattering may significantly affect the yield and statistics of the generated photons, manifested as a nonzero probability to detect a lone signal or idler whose pair partner was lost \citep{Azzini2012,Azzini2012a}. We formally account for these effects by the inclusion of $H_{bath}$, which would in principle accommodate modes into which ring photons can scatter. Since the local field intensity is strongest within the ring resonator, we assume that the effects of loss there dominate those in the channel. We therefore model the bath Hamiltonian as a set of continua of modes into which each ring mode can emit. As demonstrated in Appendix \ref{appendix:loss}, this loss mechanism is equivalent to introducing an effective ``phantom channel", identical to our physical channel, to which each ring mode linearly couples. We therefore take
\begin{eqnarray}
H_{bath} = H_{phantom\;channel} + H_{phantom\;coupling}
\end{eqnarray}
where
\begin{eqnarray}\label{phantom_channel}
&& H_{phantom\;channel} = \nonumber\\
&& \sum_J \Biggl[ \hbar\omega_J\int dz \phi_J^\dagger(z)\phi_J(z) \nonumber\\
&& + \frac{i\hbar u_J}{2}\int dz\left(\frac{d\phi_J^\dagger(z)}{dz}\phi_J(z) - \phi_J^\dagger(z)\frac{d\phi_J(z)}{d_z}\right)\Biggl]
\end{eqnarray}
and
\begin{eqnarray}\label{phantom_coupling}
H_{phantom\;coupling} = \sum_J\left(\hbar\mu_Jb_J^\dagger\phi_J(0) + \mathrm{h.c.}\right).
\end{eqnarray}
In the above expressions we have introduced effective phantom channel fields $\phi_J$, satisfying the same commutation relations as (\ref{commutation_relations}). The effective velocities $u_J$ are allowed to differ from the $v_J$, and the reference frequencies $\omega_J$ are identical to those for the ring modes. In this manner a photon lost to scattering can be thought of as existing and propagating away in the effective loss channel. When calculating physical quantities with our model we need only average over all such loss events.

\section{Equations of Motion}\label{sec:eqns_of_motion}
Having constructed our Hamiltonian, we now adopt a Heisenberg picture and compute the equations of motion for each operator. We then simplify the coupling description by introducing slowly-varying incoming and outgoing field operators, reducing the entire problem to solving a set of coupled nonlinear ordinary differential equations for the ring operators alone.
\subsection{Heisenberg equations}
Let $t_0$ be the initial time. For each Schroedinger operator $\mathcal{O}=\mathcal{O}(t_0)$ we introduce the corresponding Heisenberg-picture operator $\mathcal{O}(t)$ via
\begin{eqnarray}
\mathcal{O}(t) = U^\dagger(t,t_0) \mathcal{O}(t_0) U(t,t_0),
\end{eqnarray}
where the time evolution operator $U(t,t_0)$ satisfies $U(t_0,t_0)=1$ and 
\begin{eqnarray}
i\hbar\frac{dU(t,t_0)}{dt} = H U(t,t_0).
\end{eqnarray}
The equation of motion for $\mathcal{O}(t)$ is then
\begin{eqnarray}
\frac{d\mathcal{O}(t)}{dt} = \frac{i}{\hbar}\left[H(t),\mathcal{O}(t)\right].
\end{eqnarray}
We now proceed to calculate the equations of motion for each of the operators that appear in the Hamiltonian (\ref{hamiltonian_split_up}). For the channel fields we obtain
\begin{eqnarray}\label{channelfield_eqn}
\left(\frac{\partial}{\partial t} + v_J\frac{\partial}{\partial z} + i\omega_J \right) \psi_J(z,t) = -i\gamma_J b_J(t)\delta(z),
\end{eqnarray}
essentially a propagation equation with a driving term at $z=0$. An analogous equation holds for the phantom channel fields $\phi_J$. The ring pump operator is found to satisfy
\begin{eqnarray}\label{pump_ring_eqn}
&\bigg(&\frac{d}{dt}+ i\omega_P + 2i\eta b_P^\dagger (t) b_P(t) \bigg)b_P(t) \\ 
&=& -i\gamma_{P}^*\psi_{P}(0,t) - i\mu_{P}^*\phi_{P}(0,t) \nonumber  - 2i\lambda^*b_P^\dagger(t)b_S(t)b_I(t). \nonumber
\end{eqnarray}
For the signal and idler modes in the ring, we obtain for the Heisenberg operators
\begin{subequations}\label{signal_idler_ring_eqns}
\begin{eqnarray}
&\bigg(&\frac{d}{dt} + i\omega_S + i\zeta b_P^\dagger (t) b_P(t) \bigg)b_S(t) = \\
 &-& i\gamma_S^*\psi_S(0,t) - i\mu_S^*\phi_S(0,t) -i\lambda b_P(t)b_P(t)b_I^\dagger(t), \nonumber\\
&\bigg(&\frac{d}{dt} + i\omega_I + i\zeta b_P^\dagger (t) b_P(t) \bigg)b_I(t) = \\
 &-& i\gamma_I^*\psi_I(0,t) - i\mu_I^*\phi_I(0,t) -i\lambda b_P(t)b_P(t)b_S^\dagger(t). \nonumber
\end{eqnarray}
\end{subequations}

\subsection{Incoming and outgoing fields}
The equations of motion (\ref{channelfield_eqn}) for the channel fields can easily be formally solved, giving
\begin{eqnarray}
&&\psi_J(z,t) = \Psi_J(z,t) \nonumber \\
&&\; - \frac{i\gamma_J}{v_J}b_J\left(t-\nicefrac{z}{v_J}\right)\left[\theta(z) - \theta(z - v_Jt)\right]e^{-i\omega_Jz/v_J},
\end{eqnarray}
where $\Psi_J(z,t)$ is any solution to the homogeneous version of (\ref{channelfield_eqn}), which must take the form $\Psi_J(z,t) = e^{-i\omega_J t}f(z-v_Jt)$, with $f$ being any operator-valued function of one variable. We may substitute this expression evaluated at $z=0$ into our ring operator equations (\ref{pump_ring_eqn}) and (\ref{signal_idler_ring_eqns}). Note that as a consequence of our point-coupling model, this solution for $\psi_J(z,t)$ is discontinuous across $z=0$, a feature that survives even in classical applications of the model. It is dealt with \cite{Chak2006,Yang2007a} by introducing the one-sided limits
\begin{eqnarray}
\psi_J(0^{\pm},t) = \lim_{z\rightarrow 0^{\pm}}\psi_J(z,t)
\end{eqnarray}
and defining $\psi_J(z,t)$ at $z=0$ via
\begin{eqnarray}
\psi_J(0,t) \rightarrow \frac{1}{2}\left(\psi_J(0^-,t) + \psi_J(0^+,t)\right).
\end{eqnarray}
Using these definitions, we can relate each field at the coupling point $z=0$ to the corresponding ring operator and the field immediately left of $z=0$ (that is, the ``incoming" field),
\begin{eqnarray}\label{couplingpoint_field}
\psi_J(0,t) \rightarrow \psi_J(0^-,t) - \frac{1}{2}\frac{i\gamma_J}{v_J}b_J(t).
\end{eqnarray}
Each field operator immediately to the right of the coupling point (the ``outgoing" field) can now be put in terms of the incoming field and the corresponding ring operator:
\begin{eqnarray}\label{channelfield_transformation1}
\psi_J(0^+,t) = \psi_J(0^-,t) - \frac{i\gamma_J}{v_J}b_J(t).
\end{eqnarray}
Using (\ref{couplingpoint_field}) (and the analogous expression for the phantom channel fields $\phi_J$) in our ring operator equations (\ref{pump_ring_eqn}) and (\ref{signal_idler_ring_eqns}), we obtain
\begin{eqnarray}
\bigg(\frac{d}{dt} &+& \barGamma_P + i\omega_P + 2i\eta b_P^\dagger(t)b_P(t) \bigg)b_P(t) \nonumber \\ 
= &-& i\gamma_P^*\psi_P(0^-,t) - i\mu_P^*\phi_P(0^-,t) \nonumber \\
 &-& 2i\lambda^*b_P^\dagger(t)b_S(t)b_I(t)
\end{eqnarray}
for the pump, and
\begin{subequations}
\begin{eqnarray}
&\bigg(&\frac{d}{dt} + \barGamma_S + i\omega_S + i\zeta b_P^\dagger(t)b_P(t) \bigg)b_S(t) \\
&=& -i\gamma_S^*\psi_S(0^-,t) - i\mu_S^*\phi_S(0^-,t) - i\lambda b_P(t)b_P(t)b_I^\dagger(t), \nonumber \\
&\bigg(&\frac{d}{dt} + \barGamma_I + i\omega_I + i\zeta b_P^\dagger(t)b_P(t)\bigg)b_I(t) \\
&=& -i\gamma_I^*\psi_I(0^-,t) - i\mu_I^*\phi_S(0^-,t) - i\lambda b_P(t)b_P(t)b_S^\dagger(t) \nonumber
\end{eqnarray}
\end{subequations}
for the signal and idler. By formally solving for the channel fields and self-consistently substituting the result into the equations of motion for the ring operators, effective damping terms $\barGamma_J$ therein are introduced,
\begin{eqnarray}\label{full_gamma_defn}
\barGamma_J = \Gamma_J + M_J,
\end{eqnarray}
where
\begin{subequations}\label{individual_gamma_defn}
\begin{eqnarray}
\Gamma_J &=& \frac{|\gamma_J|^2}{2v_J}, \\
M_J &=& \frac{|\mu_J|^2}{2u_J}.
\end{eqnarray}
\end{subequations}
The $\barGamma_J$ damping constants each contain a contribution from the coupling to the physical channel, as well as a contribution from loss to the phantom channel. 

To avoid explicitly dealing with field discontinuities at the coupling point, we construct formal entities that we identify with incoming and outgoing fields. Define the incoming field $\psi_{J<}$ via
\begin{eqnarray}
\psi_{J<}(z,t) = \psi_J(z,t)\;\; \mathrm{for}\; z<0
\end{eqnarray}
and extend it to $z>0$ by demanding everywhere that
\begin{eqnarray}
\left(\frac{\partial}{\partial t} + v_J\frac{\partial}{\partial z} + i\omega_J\right)\psi_{J<}(z,t) = 0.
\end{eqnarray}
That is, we give to $\psi_{J<}$ a false future to the right of the coupling point, which corresponds to the free evolution in the absence of any coupling to the ring. We also define the outgoing field $\psi_{J>}$ via
\begin{eqnarray}
\psi_{J>}(z,t) = \psi_J(z,t)\;\; \mathrm{for}\;z>0,
\end{eqnarray}
and similarly give it a false past by enforcing everywhere
\begin{eqnarray}
\left(\frac{\partial}{\partial t} + v_J\frac{\partial}{\partial z} + i\omega_J\right)\psi_{J>}(z,t) = 0.
\end{eqnarray}
Since we will be interested in the properties of the generated photons, which exit the ring and propagate to the right, all physical quantities of interest (such as photon pair generation probabilities) will involve the outgoing channel fields $\psi_{J>}$. Indeed, since these fields follow the equation of motion corresponding to free evolution, the field at large positive $z$ (where measurements on the generated light would occur) is completely determined by the behaviour at $z=0$:
\begin{eqnarray}
\psi_J(z,t) = e^{-i\omega_Jz/v_J}\psi_{J>}(0, t - \nicefrac{z}{v_J}) \; \mathrm{for}\;z>0.
\end{eqnarray} 
Our goal is therefore to construct a solution for these outgoing fields in terms of the incoming fields $\psi_{J<}$, with which our initial pump state will be expressed. It suffices to do so at the origin, since the outgoing fields for $z>0$ are trivially related to those at $z=0$.

Another advantage of the trivial free evolution dynamics obeyed by $\psi_{J<}$ and $\psi_{J>}$ is that each field can be Fourier decomposed in a simple way:
\begin{subequations}
\begin{eqnarray}
\psi_{J<}(z,t) &&= \int \frac{d\kappa}{\sqrt{2\pi}}a_J(\kappa)e^{i\kappa z}e^{-i(\omega_J + \kappa v_J)t}, \\
\psi_{J>}(z,t) &&= \int \frac{d\kappa}{\sqrt{2\pi}}c_J(\kappa)e^{i\kappa z}e^{-i(\omega_J + \kappa v_J)t},
\end{eqnarray}
\end{subequations}
wherein the $a_J(\kappa)$ and $c_J(\kappa)$ are time-independent. These amplitudes are essentially the annihilation operators for the modes that make up the $\psi_{J<}$ and $\psi_{J>}$ fields on the respective half lines $z<0$ and $z>0$, and obey the correct bosonic commutation relations
\begin{eqnarray}
\left[a_J(\kappa),a_{J'}^\dagger(\kappa')\right] &&= \delta(\kappa - \kappa')\delta_{JJ'}, \nonumber \\
\left[c_J(\kappa),c_{J'}^\dagger(\kappa')\right] &&= \delta(\kappa - \kappa')\delta_{JJ'}.
\end{eqnarray}
Within our treatment calculations are most naturally done with respect to these annihilation operators in place of the $A_J(\kappa)$ defined in equation (\ref{original_field_defn}). We may perform the exact same procedure for the effective phantom channel fields, introducing 
\begin{subequations}
\begin{eqnarray}
\phi_{J<}(z,t) &&= \int \frac{d\kappa}{\sqrt{2\pi}}d_J(\kappa)e^{i\kappa z}e^{-i(\omega_J + \kappa u_J)t}, \\
\phi_{J>}(z,t) &&= \int \frac{d\kappa}{\sqrt{2\pi}}g_J(\kappa)e^{i\kappa z}e^{-i(\omega_J + \kappa u_J)t}.
\end{eqnarray}
\end{subequations}
The outgoing phantom channel fields $\phi_{J>}$ will contain all the photons lost to scattering or absorption. The incoming phantom channel fields will also be important, as they will amount to an extra source of vacuum fluctuations injected into the ring, which are critical for seeding the spontaneous nonlinear processes that give rise to generated signal and idler photons.

It will be convenient to remove the rapidly varying phase from the incoming and outgoing fields, as well as the ring operators, defining the full envelope quantities
\begin{eqnarray}
\overline{\mathcal{O}}_J(t) = e^{i\omega_J t}\mathcal{O}_J(t)
\end{eqnarray}
for each operator $\mathcal{O}_J(t)$. In terms of these quantities, our ring operator equations become
\begin{widetext}
\begin{subequations}\label{ring_eqns_master}
\begin{eqnarray}
\left(\frac{d}{dt} + \barGamma_P + 2i\eta\barb_P^\dagger(t)\barb_P(t)\right)\barb_P(t) &=& -i\gamma_P^*\barpsi_{P<}(0,t) - i\mu_{P}^*\barphi_{P<}(0,t) - 2i\lambda^*\barb_P^\dagger(t)\barb_S(t)\barb_I(t)e^{-i\Delta t},  \label{ring_pump_master} \\
\left(\frac{d}{dt} + \barGamma_S + i\zeta\barb_P^\dagger(t)\barb_P(t)\right)\barb_S(t) &=& -i\gamma_S^*\barpsi_{S<}(0,t) - i\mu_S^*\barphi_{S<}(0,t) -i\lambda\barb_P(t)\barb_P(t)\barb_I^\dagger(t)e^{i\Delta t}, \label{ring_signal_master}\\
\left(\frac{d}{dt} + \barGamma_I + i\zeta\barb_P^\dagger(t)\barb_P(t)\right)\barb_I(t) &=& -i\gamma_I^*\barpsi_{I<}(0,t) - i\mu_I^*\barphi_{I<}(0,t) -i\lambda\barb_P(t)\barb_P(t)\barb_S^\dagger(t)e^{i\Delta t},
\end{eqnarray}
\end{subequations}
\end{widetext}
where we have introduced the detuning parameter
\begin{eqnarray}
\Delta = \omega_S + \omega_I - 2\omega_P.
\end{eqnarray}
These ring equations, together with the transformation between the incoming and outgoing channel modes,
\begin{eqnarray}\label{incoming_outgoing_master}
\barpsi_{J>}(0,t) = \barpsi_{J<}(0,t) - \frac{i\gamma_J}{v_J}\barb_J(t),
\end{eqnarray}
constitute the fundamental equations of interest for the ring-channel system.

\section{Solving the Equations of Motion}\label{sec:solving_eqns_of_motion}
In the previous section, the problem of calculating the outgoing fields in terms of the incoming fields was reduced to solving a set (\ref{ring_eqns_master}) of coupled ordinary differential operator equations. These equations fully retain the quantum mechanical nature of all light modes, and may therefore be used to treat arbitrary injected pump states. In many experiments, however, the pump is a coherent laser beam or pulse train \cite{Azzini2012,Azzini2012a}, well described by a classical function of time. As a first example we study the semi-classical versions of equations (\ref{ring_eqns_master}), in which the pump mode $\barb_P(t)$ is replaced by its expectation value: 
\begin{eqnarray}
\barb_P(t) \rightarrow \barbeta_P(t) = \langle \barb_P(t) \rangle.
\end{eqnarray}
At this stage we also implement the undepleted pump approximation. In the ring pump equation (\ref{ring_pump_master}) the term multiplying $b_S(t)b_I(t)$ accounts for the effect on the pump mode when a photon pair is generated. Neglecting such effects, we drop this nonlinear term in the pump equation. We note, however, that pump losses are still accounted for within this approximation, as evidenced by the coupling of the pump operator to the corresponding phantom channel field. ``Undepleted" in this context therefore refers only to the nonlinear interaction; the effects of linear coupling and loss are included in our description of the pump.

With the semi-classical substitution, the (undepleted) pump equation (\ref{ring_pump_master}) becomes an entirely classical and self-contained ordinary differential equation. In the absence of SPM its solution is trivial, but in general it must be solved numerically. The signal and idler equations (\ref{ring_eqns_master}b-c) may be solved via a Green function strategy: these equations may be written in the form
\begin{eqnarray}\label{matrix_eqn_of_motion}
\frac{d}{dt}\begin{bmatrix}
\barb_S(t) \\
\barb_I^\dagger(t)
\end{bmatrix}
=M(t)\begin{bmatrix}
\barb_S(t) \\
\barb_I^\dagger(t)
\end{bmatrix}
+ D(t),
\end{eqnarray}
where we have introduced the matrix
\begin{eqnarray}
M(t) = \begin{bmatrix}
-\barGamma_S -i\zeta |\barbeta_P(t)|^2 & -i\lambda\left[\barbeta_P(t)\right]^2 e^{i\Delta t} \\
i\lambda^*\left[\barbeta_P^*(t)\right]^2 e^{-i\Delta t} & -\barGamma_I +i\zeta^* |\barbeta_P(t)|^2
\end{bmatrix}
\end{eqnarray}
as well as the driving term
\begin{eqnarray}
D(t) = \begin{bmatrix}
-i\gamma_S^*\barpsi_{S<}(0,t) - i\mu_S^*\barphi_{S<}(0,t) \\
i\gamma_I\barpsi_{I<}^\dagger(0,t) + i\mu_I\barphi_{I<}^\dagger(0,t)
\end{bmatrix}.
\end{eqnarray}
Provided a $2\times 2$ matrix $G(t,t')$ may be found which satisfies
\begin{eqnarray}\label{G_eqn_of_motion}
\frac{d}{dt}G(t,t') = M(t)G(t,t')
\end{eqnarray}
for $t>t'$ subject to initial conditions
\begin{eqnarray}
G(t',t') = \begin{bmatrix}1 & 0\\ 0 & 1\end{bmatrix},
\end{eqnarray}
then the full solution to (\ref{matrix_eqn_of_motion}) is simply
\begin{eqnarray}
\begin{bmatrix}
\barb_S(t) \\ \barb_I^\dagger(t)
\end{bmatrix}
&=& G(t,t_0)\begin{bmatrix}
\barb_S(t_0) \\ \barb_I^\dagger(t_0)
\end{bmatrix} \nonumber \\
&+& \int dt' \theta(t-t') G(t,t')D(t').
\end{eqnarray}
The signal and idler ring operator dynamics are thus easily transformed into a single $2\times 2$ matrix ordinary differential equation (\ref{G_eqn_of_motion}). Since the matrix $G(t,t')$ is complex-valued (as opposed to operator-valued), this system is easily solvable numerically. The full effect of XPM and SPM, as well as strong pumps and nontrivial pump pulse envelopes can then be investigated. We intend to carry out such a general analysis in future communications.

\section{Perturbative solution}\label{sec:perturbative}
While (\ref{G_eqn_of_motion}) must ultimately be integrated numerically, as an analytic example we now work out the weakly driven case, where we expect a perturbative approach to adequately capture the physics of photon pair generation; we neglect XPM and SPM, setting $\eta=\zeta=0$. For simplicity we also take the loss-channel propagation speeds to equal those in the physical channel, $u_J=v_J$ for each $J$, though our results (\ref{final_phi}, \ref{r_perturb}) do not depend on this assumption; they depend only on the full quantities $\Gamma_J$ and $M_J$ (\ref{full_gamma_defn}, \ref{individual_gamma_defn}).

In solving our fundamental equations the physics will be most intuitive working in the frequency domain. We define for each ring operator the corresponding Fourier amplitudes:
\begin{eqnarray}
\barb_J(t) = \int \frac{d\kappa}{\sqrt{2\pi}} \barb_J(\kappa) e^{-i\kappa v_J t}
\end{eqnarray}
so that
\begin{eqnarray} \label{fourier_relation}
\barb_J(\kappa) = v_J\int \frac{dt}{\sqrt{2\pi}} \barb_J(t) e^{i\kappa v_J t}.
\end{eqnarray}

\subsection{Channel transformations}
Consider first the (semi-classical and undepleted) pump equation (\ref{ring_pump_master}). Let our initial time be $t_0$. The homogeneous contribution to to the solution of this driven, damped, linear ordinary differential equation can be dropped by assuming there is no field in the ring at $t_0$. Remaining then is the particular solution $\barbeta_P(t)$, which in the frequency domain will satisfy
\begin{eqnarray}
\barbeta_P(\kappa) = \frac{-i\gamma_P^* \langle a_P(\kappa) \rangle}{-i\kappa v_P + \barGamma_P},
\end{eqnarray}
where we have taken $\langle d_P(\kappa) \rangle=0$, since we assume there is no incoming field in the phantom channel. This solution is exact within the semi-classical undepleted pump approximation. 

We may perform the same procedure to find $\barb_S(\kappa)$ and $\barb_I(\kappa)$ to zeroth order in the nonlinear coupling. In particular, for the idler we obtain to zeroth order
\begin{eqnarray}
\barb_I^{(0)}(\kappa) = \frac{-i\gamma_I^*a_I(\kappa) - i\mu_I^*d_I(\kappa)}{-i\kappa v_I + \barGamma_I}.
\end{eqnarray}
Inserting this and the pump solution into the frequency-domain version of our signal equation (\ref{ring_signal_master}), we find that to first order in the nonlinear coupling $\barb_S(\kappa)$ satisfies 
\begin{eqnarray}
 (-i\kappa v_S + \barGamma_S)&&\barb_S(\kappa)= -i\gamma_S^*a_S(\kappa) - i\mu_S^*d_S(\kappa) \nonumber \\
&&\; -i\lambda v_S\int d\kappa' F_S(\kappa,\kappa')\left[\barb_I^{(0)}(\kappa')\right]^\dagger,
\end{eqnarray}
where 
\begin{eqnarray}\label{F_defn}
F_S&&(\kappa,\kappa')= \nonumber \\ 
&&\bigg[ \frac{1}{2\pi}\int d\kappa_1 \int d\kappa_2 \barbeta_{P}(\kappa_1)\barbeta_{P}(\kappa_2) \nonumber \\
&& \;\;\;\;\times \delta\big(\kappa v_S - \kappa_1 v_P - \kappa_2 v_P + \kappa'v_I + \Delta)\bigg].
\end{eqnarray}
Solving for $\barb_S(\kappa)$ and substituting the result into the frequency-domain version of (\ref{incoming_outgoing_master}),
\begin{eqnarray}
c_S(\kappa) = a_S(\kappa) - \frac{i\gamma_S}{v_S}\barb_S(\kappa),
\end{eqnarray} 
we obtain for the outgoing signal field amplitudes in the channel
\begin{eqnarray}\label{perturbative_signal_transformation}
c_S(\kappa) = \int && d\kappa'\bigg[q_{SS}(\kappa,\kappa')a_S(\kappa') + p_{SS}(\kappa,\kappa') d_S(\kappa')  \nonumber \\
&& + q_{SI}(\kappa,\kappa')a_I^\dagger(\kappa') + p_{SI}(\kappa,\kappa')d_I^\dagger(\kappa')\bigg],
\end{eqnarray}
where we have introduced the spectral response functions
\begin{subequations}\label{perturbative_distribution_functions}
\begin{eqnarray}
q_{SS}(\kappa,\kappa') &&= \frac{-i\kappa v_S - \Gamma_S + M_S}{-i\kappa v_S + \barGamma_S}\delta(\kappa-\kappa'), \label{q_SS}\\
p_{SS}(\kappa,\kappa') &&= \frac{-\gamma_S\mu_S^*/v_S}{-i\kappa v_S + \barGamma_S}\delta(\kappa-\kappa'), \label{p_SS}\\
q_{SI}(\kappa,\kappa') &&= \frac{-i\gamma_S\gamma_I \lambda F_S(\kappa,\kappa')}{(-i\kappa v_S + \barGamma_S)(i\kappa' v_I + \barGamma_I)} \label{q_SI}, \\
p_{SI}(\kappa,\kappa') &&= \frac{-i\gamma_S\mu_I \lambda F_S(\kappa,\kappa')}{(-i\kappa v_S + \barGamma_S)(i\kappa' v_I + \barGamma_I)}.\label{p_SI}
\end{eqnarray}
\end{subequations}
Using an identical procedure, we may write down the channel transformation for the idler modes. We obtain
\begin{eqnarray}\label{perturbative_idler_transformation}
c_I(\kappa) = \int && d\kappa'\bigg[q_{II}(\kappa,\kappa')a_I(\kappa') + p_{II}(\kappa,\kappa') d_I(\kappa')  \nonumber \\
&& + q_{IS}(\kappa,\kappa')a_S^\dagger(\kappa') + p_{IS}(\kappa,\kappa')d_S^\dagger(\kappa')\bigg],
\end{eqnarray}
where the quantities $q_{Ix}$ and $p_{Ix}$ for $x=\lbrace S,I \rbrace$, as well as $F_I$, can be obtained by swapping the labels $S\leftrightarrow I$ in (\ref{perturbative_distribution_functions}) and (\ref{F_defn}). Analogous functions can also be derived for the transformation between the incoming fields and the outgoing loss-channel fields.

From these expressions emerges the outline of a physical picture of the system's dynamics in the regime of weak nonlinearity. Within this perturbative approximation, equations (\ref{perturbative_signal_transformation}) and (\ref{perturbative_distribution_functions}) describe the transformation between the incoming and outgoing signal field modes in the channel. This formalism is reminiscent of the conventional treatment of an optical beamsplitter, in which the outgoing and incoming mode annihilation operators are related by a simple linear transformation \cite{GerryKnight2004}. For our system there are a continuum of incoming and outgoing modes in the channel, with the ring playing the role of a ``nonlinear beamsplitter" that scatters pump photons into the signal and idler modes. As evident from the Dirac delta distribution in (\ref{q_SS}), the outgoing signal modes are correlated only with incoming signal modes of equal energy. On the other hand, from (\ref{q_SI}) we see that the outgoing signal modes are correlated with incoming idler modes having a range of energies determined by the pump-dependent function $F_S(\kappa,\kappa')$, which describes an average over all energy-conserving ways for two pump photons to be converted into a signal and idler photon pair. 

It is important to note that while this perturbative solution treated the pump semi-classically, it is straightforward to retain the pump's quantum-mechanical nature. Keeping the pump modes described quantum-mechanically allows us to treat, for example, the case of a two-photon Fock state in the pump producing a signal and idler pair. More generally, this solution can readily encompass pump states with zero expectation value for the electric field. Our semi-classical case refers to the more common laboratory setting, involving a coherent pump laser pulse, which may be treated by taking the initial state of system to be a coherent state in each pump mode, and vacuum elsewhere:
\begin{eqnarray}\label{coherent_state_defn}
\vert \psi(t_0) \rangle = e^{\int d\kappa \alpha_P(\kappa) a_P^\dagger(\kappa) - \mathrm{h.c.} } \vert \mathrm{vac} \rangle,
\end{eqnarray}
where $\alpha_P(\kappa)$ is the amplitude in mode $\kappa$ of the incoming pump field. Within this state our semi-classical pump amplitude $\barbeta_P(\kappa)$ is simply
\begin{eqnarray}\label{classical_substitution}
\barbeta_P(\kappa) = \frac{-i\gamma_P^*}{-i\kappa v_P + \barGamma_P}\alpha_P(\kappa).
\end{eqnarray} 

Also evident from the structure of our perturbative solution is how the generalization to multiple signal and idler fields would play out.  At the level of a weak undepleted pump, the presence of extra signal and idler modes would not affect the individual $q_{xx'}(\kappa,\kappa')$ and $p_{xx'}(\kappa,\kappa')$ functions. Each additional signal and idler mode would have its own set of response functions, formally identical to (\ref{perturbative_distribution_functions}), with the appropriate coupling constants and linewidths.
\subsection{Physical quantities}\label{sec:physical_quantities}
Having derived the transformations between the incoming and outgoing channel field operators in the perturbative regime, we are ready to calculate physical quantities. We will represent the initial state in the form of equation (\ref{coherent_state_defn}) as a classical, resonant Gaussian pump pulse of duration $\sigma$,
\begin{eqnarray}
\alpha_P(\kappa) = \overline{\alpha}_P e^{- (\kappa v_P \sigma / 2)^2},
\end{eqnarray}
with vacuum in all other modes.

Our focus will be on those physical quantities measurable with frequency-resolving photodetectors located at large positive $z$ in the outgoing channel. Photodetection probabilities are then given by the appropriate Glauber formula involving the expectation value of products of outgoing channel field creation/annihilation operators \cite{GerryKnight2004}. The probability of detecting a signal and idler photon pair with wavevectors offset from the reference wavevectors $\kappa_S$ and $\kappa_I$ by $\kappa$ and $\kappa'$, respectively, is proportional to
\begin{eqnarray}
\Phi(\kappa,\kappa') = \langle c_S^\dagger(\kappa) c_I^\dagger(\kappa') c_S(\kappa) c_I(\kappa')  \rangle,
\end{eqnarray}
where the expectation value is taken with the state $\vert \mathrm{vac} \rangle$, the vacuum of the incoming channel field operators. This is exactly the joint spectral intensity of the signal and idler fields. Using our perturbative response functions (\ref{perturbative_distribution_functions}) we obtain (to leading order in $\lambda$, in keeping with the order of the perturbative solution)
\begin{eqnarray}\label{final_phi}
\Phi(\kappa,\kappa') = \frac{|\lambda|^2|\gamma_S|^2|\gamma_I|^2|F_S(\kappa,\kappa')|^2}{((\kappa v_S)^2 + \barGamma_S^2) ( (\kappa' v_I)^2 + \barGamma_I^2)}.
\end{eqnarray}
The shape of this function is determined by the pump through $F_S(\kappa,\kappa')$ -- in particular, the pump pulse bandwidth determines the width of this two-photon distribution, as well as the degree of spectral correlation between the two photons in the pair. 

\begin{figure}
\includegraphics[width=1.0\columnwidth]{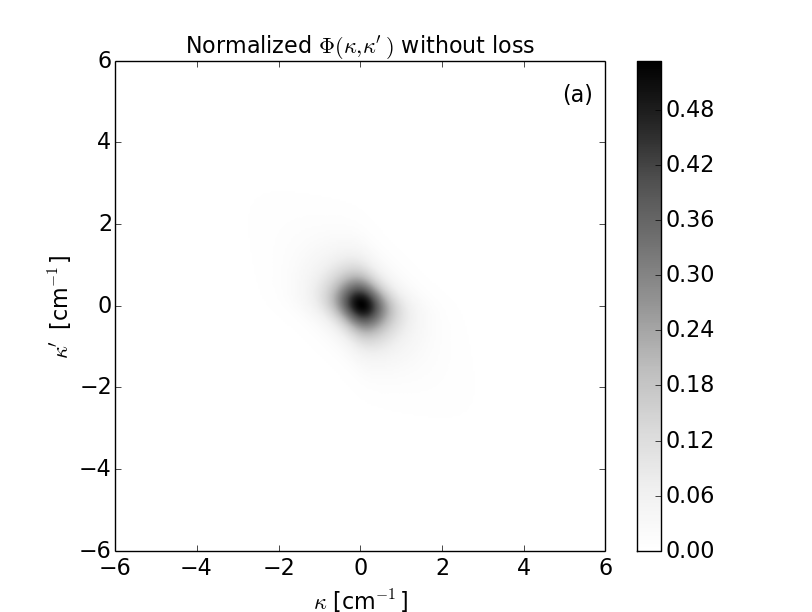}
\includegraphics[width=1.0\columnwidth]{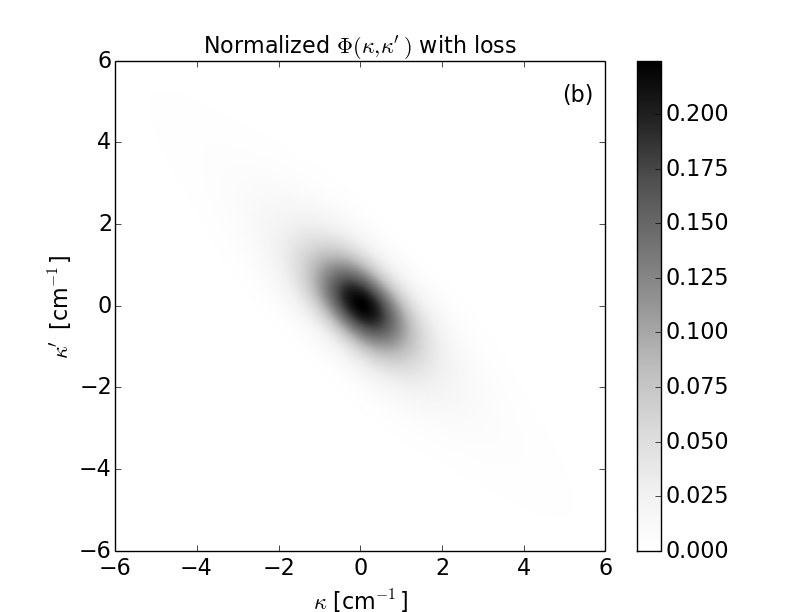}
\caption{Normalized joint spectral intensity function for a 100 ps pulse incident on a ring (a) without loss, $\barGamma_J=10$ GHz and (b) with loss, $\barGamma_J=20$ GHz. Propagation speeds for all modes are taken to be $v_J=15$ cm/ns.\label{fig:phi_plot}}
\end{figure}

\begin{figure}
\includegraphics[width=1.0\columnwidth]{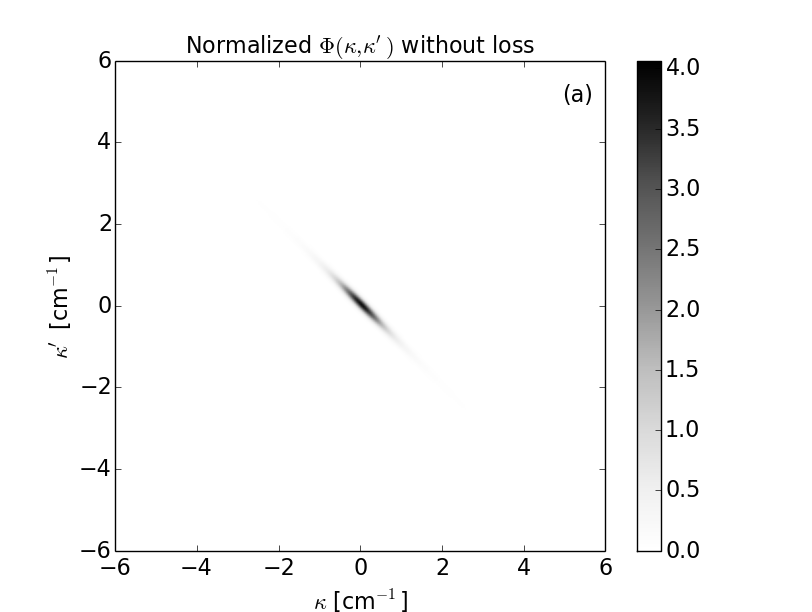}
\includegraphics[width=1.0\columnwidth]{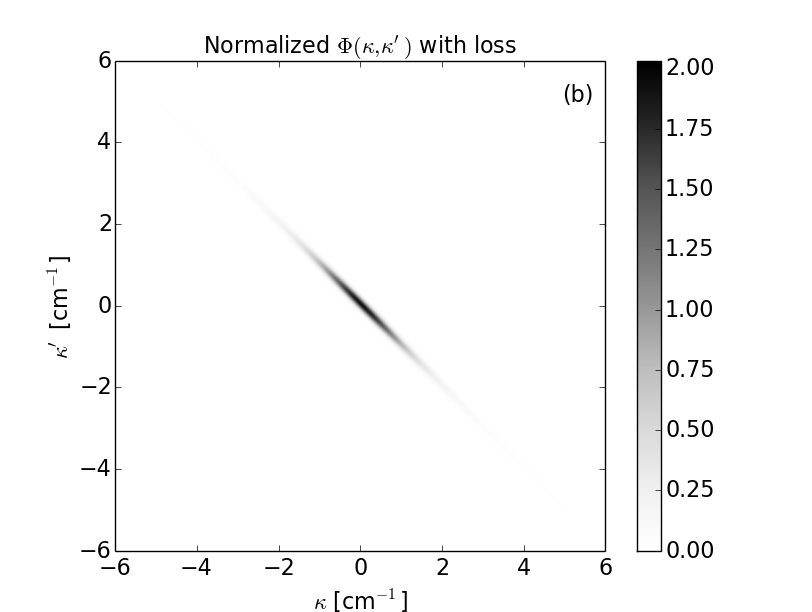}
\caption{Normalized joint spectral intensity function for a 1 ns pulse incident on a ring (a) without loss, $\barGamma_J=10$ GHz and (b) with loss, $\barGamma_J=20$ GHz. Propagation speeds for all modes are taken to be $v_J=15$ cm/ns.\label{fig:phi_plot2}}
\end{figure}

Also evident from this expression is the effect of loss on $\Phi(\kappa,\kappa')$. Equation (\ref{final_phi}) is in formal agreement with earlier work \cite{Helt2010} where loss was neglected, but with $\Gamma_J$ now replaced by $\barGamma_J$. That is, the inclusion of loss is completely encapsulated in the total effective damping rates $\barGamma_S$ and $\barGamma_I$, which affect the widths of the overall Lorentzian envelopes in the denominator, as well as the transfer of pump energy from the channel to the ring as described by equation (\ref{classical_substitution}). While the phantom channel fields do contribute to the vacuum fluctuations that seed the pair generation process, as evidenced by the presence of the $p_{xx'}$ functions in (\ref{perturbative_signal_transformation}), the overall yield of pairs that exit to the physical channel only depends on loss insofar as the ring linewidths are affected. Plots of the normalized joint spectral intensities for a 100 ps pulse incident on a ring without and with loss ($\barGamma_J=10$ GHz and $\barGamma_J=20$ GHz, respectively) in the regime where the pulse duration is comparable to the lossless ring photon lifetimes are illustrated in Fig. (\ref{fig:phi_plot}). In this regime, in the absence of loss, the signal and idler energies are nearly uncorrelated, as evidenced by the circular shape of the joint spectral intensity. The inclusion of the phantom channel with coupling strengths equal to those for the physical channel doubles the total effective damping rate, leading to significant broadening of the overall distribution, and introduces some degree of spectral correlation. Fig. (\ref{fig:phi_plot2}) illustrates the opposite regime, in which a longer 1 ns pulse is employed to yield entanglement between the signal and idler photons. Loss similarly leads to broadening of the overall signal and idler linewidths, but does not alter the degree of spectral correlation, indicated by the breadth of the distribution in the direction orthogonal to the antidiagonal.

Though loss can be fully accounted for in the joint spectral intensity as corrections to the ring linewidths, the inclusion of the phantom channel introduces important quantum effects in the statistics of the outgoing photons in the physical channel. These arise from the probability for one photon in a generated pair to be lost to scattering in the ring, with only its partner subsequently available for detection. The total rate of pair detection events -- the ``coincidences rate" -- is simply proportional to
\begin{eqnarray}
P_{coincidences} = \int d\kappa \int d\kappa' \Phi(\kappa,\kappa').
\end{eqnarray}
This quantity must be distinguished from the total \emph{singles} probability,
\begin{eqnarray}
P_{singles} = \int d\kappa \int d\kappa' \bigg[&& \langle c_S^\dagger(\kappa)g_I^\dagger(\kappa')c_S(\kappa)g_I(\kappa')\rangle \\
&& + \langle c_I^\dagger(\kappa)g_S^\dagger(\kappa')c_I(\kappa)g_S(\kappa')\rangle \bigg] \nonumber,
\end{eqnarray}
which represents the detection of a signal or idler photon whose entangled partner was lost to scattering in the ring. Of particular interest is the ratio between this quantity (the ``singles rate") and the coincidences rate,
\begin{eqnarray}
r = \frac{P_{singles}}{P_{coincidences}}.
\end{eqnarray}
The inclusion of a loss mechanism in our treatment allows us to calculate this quantity, which in previous theoretical treatments would always equal zero. A more realistic estimate of $r$ is especially important to applications wherein one of the photons in a pair is used to trigger another branch of the experiment that requires the trigger photon's pair partner. Actuating a detector off of a photon whose partner was lost would typically be undesirable, making it necessary to understand the origin and magnitude of this effect. In the perturbative regime, we find
\begin{eqnarray} \label{r_perturb}
r = \frac{\Gamma_S M_I + \Gamma_I M_S}{\Gamma_S \Gamma_I}.
\end{eqnarray}
This expression is entirely independent of the pump and nonlinear coupling strength. In actual experimental implementations, the ring-channel system is often operated at or near the critical coupling regime, where the linear transmissivity through the device is minimized \cite{Azzini2012,Azzini2012a}. Within our treatment this occurs when the coupling rates between the ring and the phantom channel equal those between the ring and the physical channel: $\Gamma_J=M_J$ for each $J$. In this regime (\ref{r_perturb}) gives $r=2$. Other losses present in an actual experiment can be expected only to increase this value, so $r=2$ can be taken as the theoretical minimum for lossy, critically coupled ring resonator systems.

\section{Conclusion}
We have developed a general Hamiltonian treatment of spontaneous four-wave mixing in a microring resonator side-coupled to a channel waveguide. The inclusion of losses in the ring permitted the calculation of the singles-to-coincidences ratio $r$, for which a lower bound of $r=2$ was derived in the critically coupled weak pumping regime in the absence of other losses. In this regime the inclusion of loss was found to affect the joint spectral intensity distribution of generated photon pairs strictly through corrections to the linewidths of the ring resonances. As such, when the pump pulse duration is comparable to the lossless ring photon lifetimes, the addition of loss introduces significant broadening and some spectral correlation to the generated photon pairs. For longer pump pulses loss merely broadens the overall joint spectral intensity, but does not appreciably change the degree of entanglement. In the strongly pumped regime, wherein self- and cross-phase modulation may become important, a Green function procedure for numerically deriving relevant physical quantities was presented. We intend to present further analyses of this regime in future works.

\appendix

\section{An equivalent channel model for loss}\label{appendix:loss}
To model scattering losses in the ring, we imagine coupling each ring mode $J$ to a continuum of bath modes described by annihilation operators $D_J(\kappa)$. The bath Hamiltonian, including the coupling to the ring modes, would be of the form
\begin{eqnarray}
H_{bath} &=& \sum_J\int d\kappa \hbar\omega_J(\kappa) D_J^\dagger(\kappa) D_J(\kappa) \nonumber \\
 &+& \sum_J\left(b_J^\dagger\int d\kappa \hbar\tilde{\mu}_J(\kappa)D_J(\kappa) + h.c.\right)
\end{eqnarray}
where $\omega_J(\kappa)$ are the bath mode frequencies, and $\tilde{\mu}_J(\kappa)$ are coupling coefficients to the corresponding ring modes. We now introduce a fictitious position parameter $z$, allowing us to define field operators $\phi_J(z)$ in the usual way,
\begin{eqnarray}
\phi_J(z) = \int d\kappa e^{i(\kappa - \kappa_{J})z} D_J(\kappa),
\end{eqnarray}
where $\kappa_{J}$ is the reference wavevector for the mode $J$. Provided the function $\tilde{\mu}_J(\kappa)$ is sufficiently smooth and slowly varying over the relevant range of $\kappa$, we can take
\begin{eqnarray}
\int d\kappa \hbar\tilde{\mu}_J(\kappa) e^{i(\kappa - \kappa_{J})z} \approx \hbar \mu_J \delta(z),
\end{eqnarray}
where $\mu_J$ is a constant. The coupling term in $H_{bath}$ then becomes
\begin{eqnarray}
&&\sum_J\left(b_J^\dagger\int d\kappa \hbar\tilde{\mu}_J(\kappa)D_J(\kappa) + h.c.\right) \nonumber \\
&\approx& \sum_J\left(\hbar\mu_J b_J^\dagger \phi_J(0) + h.c.\right).
\end{eqnarray}

If we now expand $\omega_J(\kappa)$ to first order in $\kappa$ about the the appropriate reference frequency,
\begin{eqnarray}
\omega_J(\kappa) = \omega_J(\kappa_{J}) + (\kappa-\kappa_{J})u_J,
\end{eqnarray}
where $u_J$ is the effective propagation speed in the phantom channel,
\begin{eqnarray}
u_J = \left(\frac{d\omega_J}{d\kappa}\right)_{\kappa_{J}},
\end{eqnarray}
then the full bath Hamiltonian becomes
\begin{eqnarray}
H_{bath} = H_{phantom\;channel} + H_{phantom\;coupling},
\end{eqnarray}
as defined in (\ref{phantom_channel}) and (\ref{phantom_coupling}).
\begin{acknowledgments}
We acknowledge support from the Natural Sciences and Engineering Research Council of Canada. Thanks are also due to Nicolas Quesada for helpful discussions.
\end{acknowledgments}

\bibliography{squeezing}

\end{document}